\documentclass[twocolumn,showpacs,preprintnumbers,amsmath,amssymb]{revtex4}
\usepackage{graphicx}
\usepackage{dcolumn}
\usepackage{bm}
\begin{document}

\widetext

\title{Low temperature transition to a superconducting phase in boron-doped
    silicon films grown on (001)-oriented silicon wafers}
\author{C. Marcenat$^1$, J. Ka\v{c}mar\v{c}\'ik$^{1,2}$, R. Piquerel$^1$, P. Achatz$^1$, G. Prudon$^3$, C. Dubois$^3$, B. Gautier$^3$, J.C. Dupuy$^3$,  E. Bustarret$^4$, L. Ortega$^4$, T. Klein,$^{4,5}$,  J. Boulmer$^{6,7}$, T.  Kociniewski$^{6,7}$, and D. D\'ebarre$^{6,7}$,}
 \affiliation{$^1$CEA-Grenoble, Institut Nanosciences et Cryog\'enie, SPSMS-LATEQS, 17 rue des Martyrs, 38054 Grenoble Cedex 9, France}
\affiliation{$^2$ Centre of Very Low Temperature Physics, IEP Slovak Academy of Sciences and FS UPJ\v{S}, 040 01 Ko\v{s}ice, Slovakia}
\affiliation{$^3$ Institut des Nanotechnologies de Lyon, CNRS and INSA, 7 av. J. Capelle, 69621 Villeurbanne Cedex, France}
\affiliation{$^4$ Institut N\'eel, CNRS, B.P. 166, 38042 Grenoble Cedex 9, France}
\affiliation{$^5$ Institut Universitaire de France and Universit\'e Joseph Fourier, B.P.53, 38041 Grenoble Cedex 9, France}
\affiliation{$^6$ Institut d'Electronique Fondamentale, Universit\'e Paris Sud, 91405 Orsay, France}
\affiliation{$^7$ Institut d'Electronique Fondamentale, CNRS, 91405 Orsay, France}

\date{\today}

\begin{abstract}
We report on a detailed analysis of the superconducting properties of boron-doped silicon films grown along the 001 direction by Gas Immersion Laser Doping. The doping concentration c$_B$ has been varied up to $\sim 10$ at.\%  by increasing the number of laser shots  to 500. No superconductivity could be observed down to 40mK for doping level below $\sim 2.5$ at.\%. The critical temperature $T_c$ then increased steeply to reach $\sim 0.6$K for c$_B \sim 8$ at\%. No hysteresis was found for the transitions in magnetic field, which is characteristic of a type II superconductor. The corresponding upper critical field $H_{c2}(0)$  was on the order of $1000$ G, much smaller than the value previously reported by Bustarret {\it et al.} \cite{Bustarret06} 
\end{abstract}

\maketitle
 
The discovery of a superconducting transition around $40$ K in the MgB$_2$ compound \cite{Nagamatsu01}  revived the interest for a specific class of superconducting materials : the covalent metals \cite{Blase09,Crespi03}. Indeed, the high $T_c$ value is here a direct consequence of the strong coupling of the boron $\sigma-$bands (shifted up to the Fermi level due to the presence of Mg$^{2+}$ ions) with phonons. This system was the precursor of new covalent superconductors among which boron doped diamond C:B \cite{Ekimov04}, silicon Si:B \cite{Bustarret06}, silicon carbide SiC:B \cite{Ren07}, as well as gallium doped germanium \cite{Hermannsdorfer09} .  

Eventhough these systems share the similarity of being heavily doped semiconductors for which superconductivity appears in the partially unfilled valence band, they also present striking differences. In diamond,  the onset of doping-induced superconductivity  coincides with the metal-insulator transition (MIT)  \cite{Klein07}. On the other hand, although silicon becomes metallic for boron doping on the order of  80 p.p.m., a concentration of several per cent was necessary to trigger superconductivity \cite{Bustarret06}.  Such a concentration  exceeds the solubility limit and can only be reached using sophisticated out-of-equilibrium doping techniques.   This  explains why a systematic study of the evolution of $T_c$ vs boron doping in Silicon is still lacking.

We present here such a study for a series of high quality boron-doped silicon films. No superconductivity could be observed down to 40mK for doping levels up to $\sim 2.5$ at.\%. $T_c$ then rises sharply with boron concentration reaching $\sim 0.6$K for c$_B \sim 8$ at.\%,  a twice larger temperature than reported previously \cite{Bustarret06}.  A similar rapid increase in $T_c$ vs doping has been reported in B doped diamond films for which $T_c$ scales as (c$_B$/c$_c$-1)$^{0.5}$  \cite{Klein07}, where c$_c$ is the critical concentration corresponding to both the MIT and the onset of superconductivity. As shown below a similar dependence may also roughly apply to the present results but would require the use of a c$_c$ value on the order of 2.5 at.\%, i.e. well above the MIT doping threshold. We also show that, despite their  higher $T_c$ values, the upper critical field $H_{c2}(0)$ of our high quality films  is about four times lower than the one previously reported  in \cite{Bustarret06}.

In order to overcome the solubility limit, a set of  samples have been prepared by Gas Immersion Laser Doping (GILD) \cite{Bustarret06,GILD}. A precursor gas (BCl$_3$) is chemisorbed on a (001)-oriented silicon wafer which is subsequently melted using an ultraviolet laser pulse. During each melting/solidification cycle, boron diffuses from the surface into molten silicon, and is incorporated at substitutional sites of the crystal as the liquid/solid interface moves back to the surface of the epilayer upon cooling after the laser pulse.  The in-plane lattice match to the substrate and the local incorporation of boron atoms of lower covalent radius  result in the formation of biaxially strained pseudo-morphic epilayers. The B concentration was progressively increased by increasing the number of laser shots, from 1.4 $at. \%$ (50 shots) up to  $\sim$10 $at.\%$ (500 shots), the laser energy being regulated in order to keep the doped layer thickness constant during the whole process, here on the order of 80-90 nm. 
Since our previous  publication \cite{Bustarret06}, new laser optics have strongly improved the spatial homogeneity in energy within the spot and an ultrahigh residual vacuum ($p  \sim 10^{-7}$ Pa) is reached in the reaction chamber.

\begin{figure}[!t]
\begin{center}
\resizebox{0.48\textwidth}{!}{\includegraphics{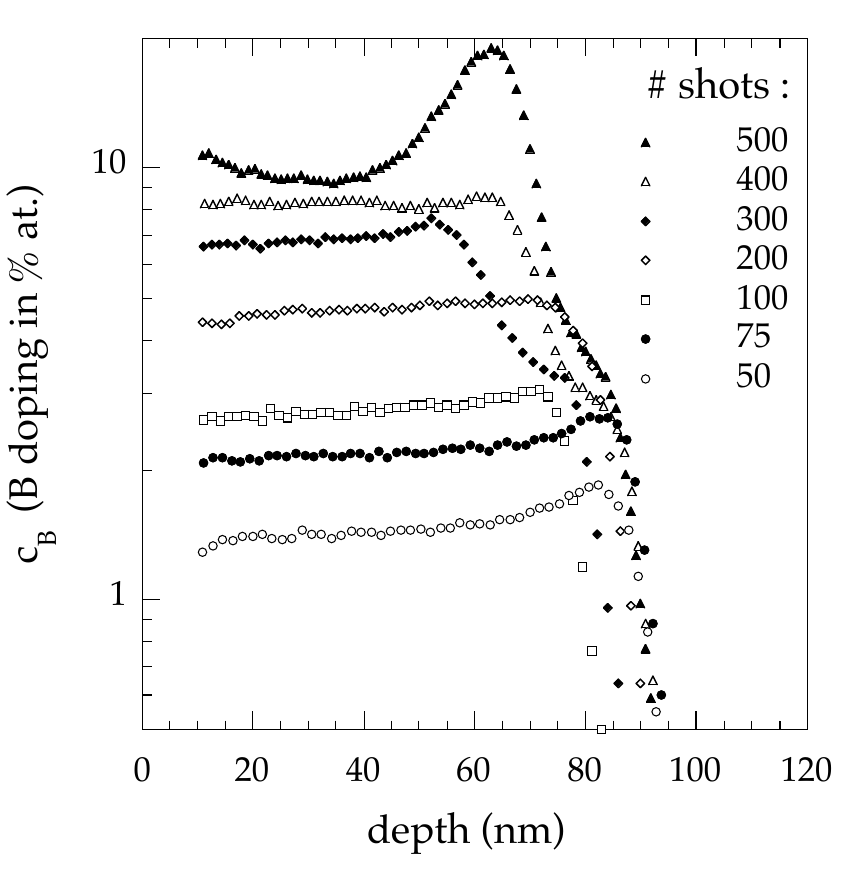}}
\caption{Boron atomic concentration depth profiles deduced from SIMS measurements for the indicated numbers of lasers shots in the GILD process. }
\label{Fig.1}
\end{center}
\end{figure}


\begin{figure}
\begin{center}
\resizebox{0.48\textwidth}{!}{\includegraphics{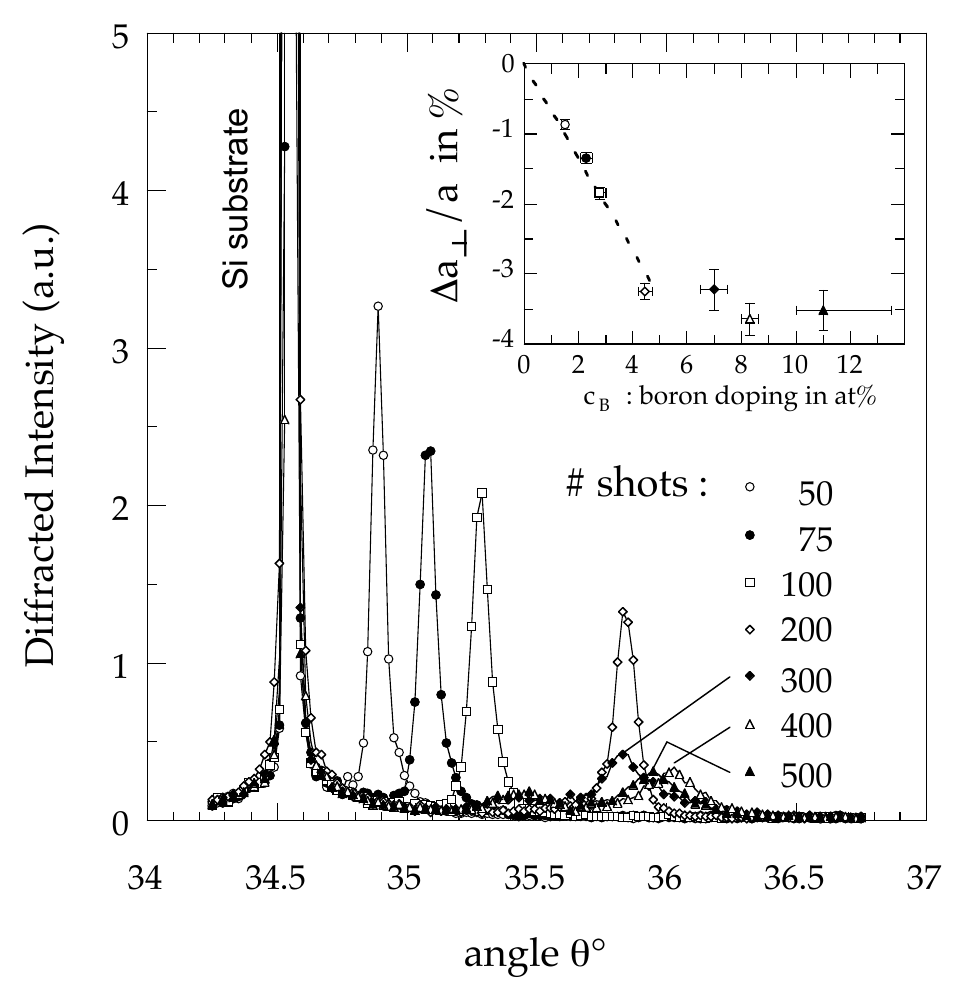}}
\caption{ High-resolution XRD measurements of Si:B films at [004] Bragg
reflection. The peak at $34.565^\circ$ corresponds to the Si substrate and the second peak is characteristic of the doped layer for the indicated numbers of laser shots in GILD process. Inset: Relative variation of the lattice parameter perpendicular to the surface in \% as a function of the boron concentration c$_B$. Broken line, linear dependence (Vegard's law) at low doping. }
\label{Fig.2}
\end{center}
\end{figure}


Measurements of boron concentration depth profiles were done by Secondary Ion Mass Spectrometry (SIMS) using a CAMECA IMS4f instrument. Reliable quantification of boron in silicon for such high concentration levels  above 1 $at. \%$ is a major challenge for SIMS analysis due to matrix effects \cite{SIMS} as complex ionization and/or sputtering mechanisms may occur during secondary ion emission. We have shown that these effects remain small with an O$_2^+$ beam in presence of a saturated oxygen flow at the surface \cite{sims O2}. A primary 4keV O$_2^+$sputtering beam of 90nA is rastered at oblique incidence (46.8$^{\circ}$) on 250x250 $\mu$m$^2$ of the surface. To avoid crater edge effects,  $^{10}$B$^+$, $^{11}$B$^+$, and  $^{29}$Si$^+$ ions coming from a central part of 10$\mu$m in diameter have been selected and analyzed simultaneously by an original Isotopic Comparative Method (ICM) (for details see \cite{ICM}). Atomic concentrations are presented in Fig. 1 as a function of the depth measured by mechanical profilometry with a Tencor P10. High resolution X-ray diffraction (XRD) curves were then collected around the [004] symmetric Bragg reflection of silicon and are displayed in Fig. 2. A monochromatized $CuK_{\alpha 1}$ excitation source led to an angular resolution better than $5$x$10^{-3}$ degree.

Data shown in Figures 1 and 2 illustrate the significant improvement of the structural quality and sample homogeneity with respect to previous publications  \cite{Bustarret06,GILD}. Up to 200 shots (c$_B \leq$ 4.5 $at.\%$), we observe relatively flat boron concentrations (Fig. 1) with a sharp interface, at 80 to 90 nm, on the order of 9 nm/decade. The peaks in XRD (Fig. 2) remain very well defined with a full width at half maximum on the order of 0.1$^{\circ}$. As the number of laser shots is increased, the shift to higher diffraction angles illustrates
the contraction of the out-of-plane lattice parameter a$_{\perp}$ as a result of the presence of boron atoms with a smaller covalent radius than silicon.
This is a good indication of a sufficiently rapid re-crystallization limiting the diffusion processes and leading to an homogeneous incorporation of boron on substitutional sites far above the solubility limit of  1.2 $at\%$. In this range, as illustrated by the inset of Fig. 2, the out-of-plane negative strain deduced from the diffraction patterns varied linearly with boron incorporation as measured by SIMS, the slope  being in fair agreement with the strain rate coefficient of Vegard's law in silicon \cite{Vailionis99} . Above 200 shots, the interface began to exhibit a shoulder or even a large peak (500 shots) indicating a pileup of boron at the interface during the process. The diffraction peaks did not shift upwards  anymore and remained close to 36$^{\circ}$. Their intensity weakened and their width increased substantially, a second weaker broad peak even developing around 35.5$^{\circ}$ for 500 laser shots. This corresponds to a saturation of the out-of-plane strain around -3.5\%. Since according to SIMS measurements the intake of boron went on increasing between 200 and 500 shots,  one is led to conclude that in this range an increasing fraction of these boron atoms were not incorporated  on substitutional sites. In our experience, the saturation value of 36.1$^{\circ}$ observed here for the [004] diffraction peak of Si:B is a quite general upper limit, independent of our  GILD operation conditions.

The superconducting temperatures and upper critical field have been deduced from ac-susceptibility ($\chi_{ac}$) and magneto-transport ($R$) measurements. In the $\chi_{ac}$ measurements, the films have been placed on top of miniature coils and $T_c$ has been obtained by detecting the change
in the self induction of the coils induced by the superconducting
transitions. Both AC and DC magnetic fields were applied perpendicularly to the doped layer. Approximate corrections were made to account for the demagnetization factor but they don't affect the determination of $H_{c2}$ where the magnetization is zero. As shown in the inset of Fig. 3, the onset of the diamagnetic response well coincides with the resistive transition, which is remarkably sharp in this sample, $\Delta T_c \sim$ 10mK, while being only  5 times larger in the other  samples of the series. This allows an accurate determination of $T_c$ and the corresponding values have been reported in Fig. 3 as a function of the B content deduced from SIMS measurements.

\begin{figure}[!t]
\begin{center}
\includegraphics[width=0.48\textwidth]{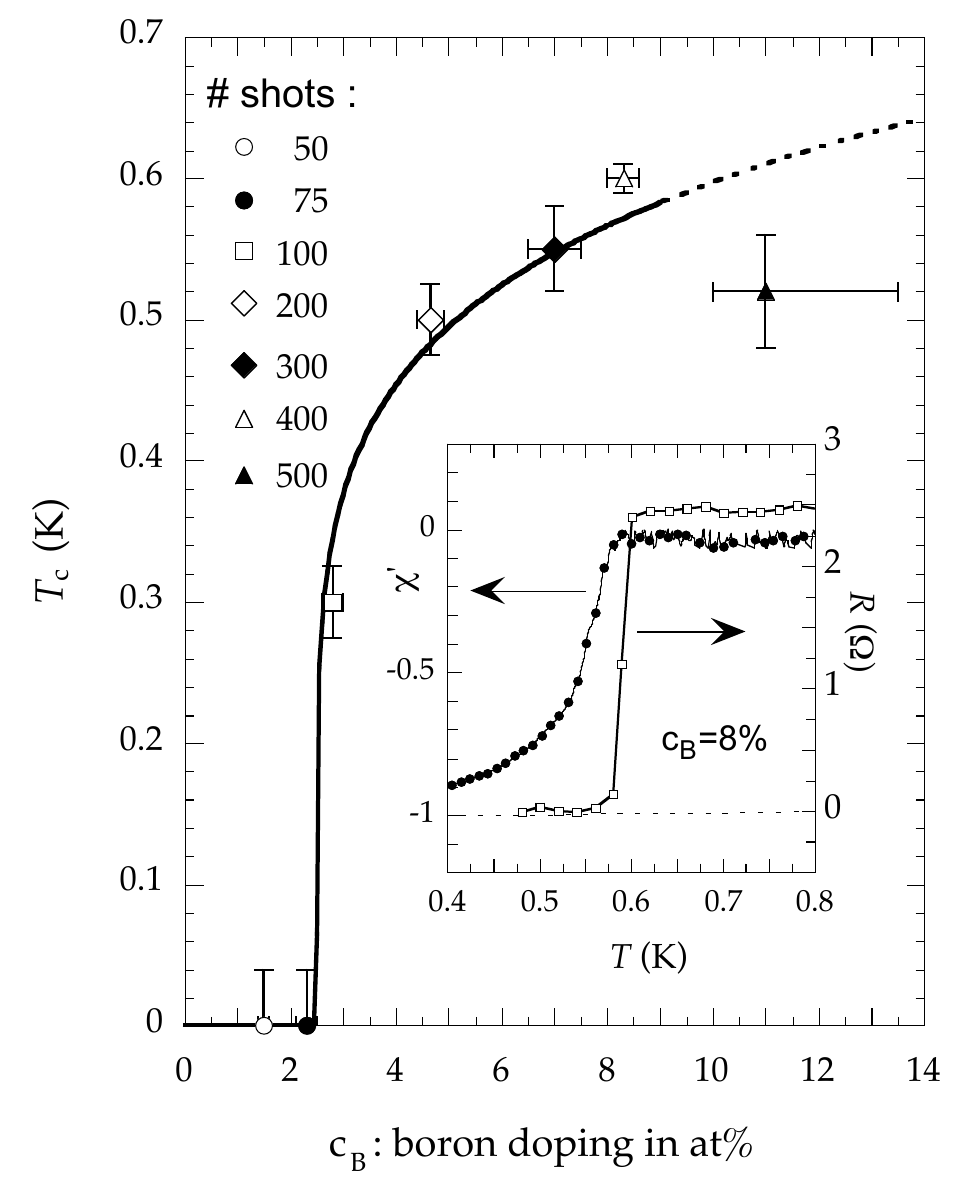}
\caption{Dependence of the superconducting transition temperature $T_c$ on the boron concentration c$_B$ deduced from SIMS measurements. In the inset :  Resistive (right scale) and magnetic (left scale) transition in zero magnetic field in a Si:B film (400 laser shots and c$_B \sim 8$ at.\%) showing that both transport and diamagnetic transitions are very sharp and occur at the same temperature.}
\label{Fig.3}
\end{center}
\end{figure}


\begin{figure}[!t]
\begin{center}
\resizebox{0.48\textwidth}{!}{\includegraphics{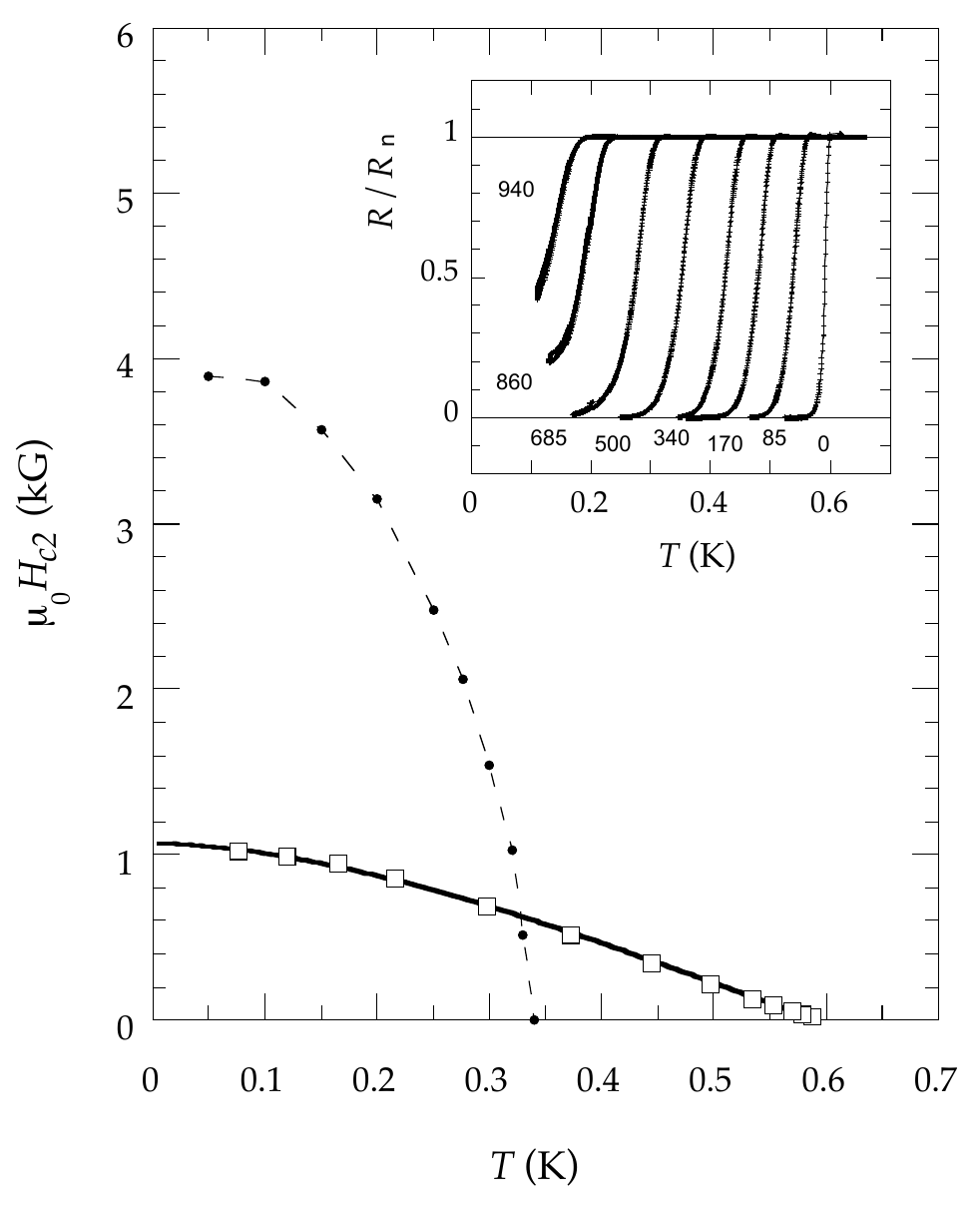}}
\caption{Temperature dependence of the upper critical field $H_{c2}$ deduced from transport measurements (see inset) in a heavily doped Si:B samples (c$_B \sim 8$ at.\%, open squares; classical theory in the very dirty limit, full line) together with the values previously reported in \cite{Bustarret06}  (dots and dashed line). In the inset: temperature dependence of the resistance normalized to its normal state value in a serie of magnetic fields given in Gauss}
\label{Fig.4}

\end{center}
\end{figure}


No superconductivity was found down to 40mK for the two less doped samples with c$_B \sim$ 1.4 $at.\%$ (50 laser shots) and c$_B \sim$ 2.2 $at.\%$ (75 laser shots). For higher boron concentration, the critical temperature increases sharply with doping above some critical concentration c$_c$ $\sim$ 2.5 $at.\%$ reaching 0.6K for c$_B$ $\sim$ 8 $at.\%$. For comparison, the sample measured in \cite{Bustarret06} displayed a roughly twice smaller $T_c$ with a much broader transition. The apparent saturation of $T_c$ is probably non-intrinsic since the incorporated boron in this range of concentration might be non-electrically active as suggested by XRD and by the observation of a maximum angle of diffraction with broad and less intense  peaks (see above). Calculations within Density Functional Theory and Virtual Crystal Approximation \cite{Boeri04} predict $T_c$ $\sim$ 0.3K (resp. 3K) for c$_B = $ 5 $at.\%$ (resp. 10 $at.\%$) using MacMillan formula \cite{Carbotte}. Even if the order of magnitude is correct, the observed dependence of $T_c$ on c$_B$ is strikingly different from these predictions. $T_c$ (c$_B$) is actually very similar to that observed in doped diamond, with a critical threshold for doping concentration c$_c$ above which $T_c$ increases very sharply with a quasi power law dependence, $T_c$ $\propto$ ($\frac{c_B}{c_c}-1)^{0.5}$. Whereas in doped diamond c$_c$ coincides with the critical concentration of the metal-insulator transition, in doped silicon c$_c$ $\sim$ 2.5 $at.\%$, i.e. about 300 times  larger than the concentration necessary to induce metallicity. The meaning of c$_c$  in this latter case remains therefore rather puzzling and needs to be clarified by further studies.

Electrical measurements were performed in magnetic field on the sample with c$_B \sim 8$ at.\% (400 laser shots) which has the highest $T_c$ and the sharpest transition in zero field. The results
are displayed in the inset of Fig. 4 as temperature was swept at different fixed magnetic fields. The transitions was shifted towards lower temperature as the magnetic field was increased, while remaining sharp and well defined. No hysteresis nor supercooling was observed, indicating a second-order transition, which is consistent with doped silicon being a type II superconductor. The  upper critical magnetic field $H_{c2}$, defined by the usual criterion as where the resistance is 90\% of its normal state value, is plotted in Fig. 4. Note that the discussion below and the shape of the critical line don't depend on the exact criterion used; a $R/R_n = $50\% definition would only shift slightly the $H_{c2}(T)$ line.

The temperature dependence of $H_{c2}$ is very well reproduced  within the standard microscopic theory. The full line in Fig. 4 is a fit to the solution of the linearized Gor'kov equations neglecting spin effects \cite{HW66} in the very dirty limit, $\frac{\ell}{\xi} \rightarrow 0 $, $\ell$  being the electronic mean-free path and $\xi$ the superconducting coherence length. Both the strong curvature and a 4 times larger $H_{c2}$($T$=0K) observed previously \cite{Bustarret06} were most probably a consequence of the film inhomogeneity. Since the Fermi velocity $v_F$ is a parameter which cannot be determined experimentally with any accuracy,  the electronic mean-free path $\ell$ and $H_{c2}(0)$ were estimated in a simple two-band free electron model \cite{Walte04} with light (subscript $lh$) and heavy (subscript $hh$) holes  having an effective mass m$_{lh}^*$ = 0.16 m$_e$ (resp. m$_{hh}^*$ = 0.49 m$_e$), m$_e$ being the bare electron mass, and with a total carrier concentration of $4$x$10^{21}$cm$^{-3}$. The residual resistivity in this sample was equal to $\rho_o$ $\sim$ 100$\mu \Omega$cm and $\ell~=~ \frac{\sqrt{<v_F^2>}} {\rho_o e^2} (\frac{m_{lh}}{n_{lh}}+\frac{m_{hh}}{n_{hh}})$ $~\sim~3nm$. The light hole band is expected to play a minor role for the determination of the upper critical field due to its larger Fermi velocity and its lower partial density of states.  In the dirty limit, $H_{c2}$  is therefore related to the electronic mean-free path $\ell$ and to the coherence length in the heavy hole band $\xi_{hh}$  by:   $H_{c2}~ \approx ~ \frac{3 \Phi_o}{2 \pi^2 l  \xi_{hh}}$ \cite{deGennes,Maki} $\sim$ 900 Gauss with  $\xi_{hh}$ being renormalized by the electron-phonon coupling constant, $\lambda_{ep} \sim 0.3$ \cite{Bustarret06} and $\xi_{hh} ~ =~ \frac{\hbar v_F^{hh}}{\pi \Delta (1+\lambda_{ep}  )}~ \sim~ 1200nm$ with $\Delta$ being the superconducting gap. Considering the uncertainties in the thickness of the film and in the geometry of the contacts in a van der Pauw configuration together with the crudeness of a free electron approximation, this estimate  $H_{c2} (0)~ \sim ~ 900$ Gauss is in good agreement with the experimental data $H_{c2} (0)~ \sim ~ 1050$ Gauss (see Fig. 4). Note that the London penetration depth $\lambda_L$ = $\sqrt{\frac{(1+\lambda_{ep} )m^*}{\mu_o n e^2}}$ $\sim$ 60nm $\leq$ $\xi$ implying that doped Si should be intrinsically a type I superconductor such as SiC:B \cite{Kriener08} but is turned into a type II system by strong impurity effects ($\kappa \sim 0.7 \lambda_L / \ell \ge$ 1 in the dirty limit but $\kappa =  \lambda_L / \xi \le$ 1 in the clean limit).

To conclude, the GILD technique is proved to be a powerful technique to dope silicon in the alloying range where superconductivity occurs. The superconducting transitions are sharp and well defined both in resistivity and magnetic susceptibility allowing the study of the variation of $T_c$ on the boron concentration. This variation is in contradiction with a classical exponential dependence on superconducting parameters. Instead, $T_c$ increases with a quasi-power law form above a critical threshold c$_c$ in striking similarity  to what is observed in doped diamond. Although in the diamond case c$_c$ was equal to the MIT critical concentration, in silicon it lies far above. The comparison of these two cases offers a new playground for studying superconductivity in covalent systems near a MIT.

\end{document}